\documentclass[]{interact}

\usepackage{epstopdf}
\usepackage{subfigure}
\usepackage[colorlinks,linkcolor=black,anchorcolor=black,citecolor=blue]{hyperref}
\usepackage{natbib}
\bibpunct[, ]{(}{)}{;}{a}{}{,}

\theoremstyle{plain}
\newtheorem{theorem}{Theorem}[section]

\newtheorem{corollary}[theorem]{Corollary}

\theoremstyle{definition}

\newtheorem{example}[theorem]{Example}

\theoremstyle{remark}
\newtheorem{remark}{Remark}

\usepackage{multirow}

\begin{document}

\articletype{}

\title{The CUSUM Test with Observation-Adjusted Control Limits in Parameters Change Detection for the Extremely Heavy-Tailed Distributions Sequences\textsuperscript{*}}

\author{
\name{Fuquan Tang\textsuperscript{a} and Dong Han\textsuperscript{b}\thanks{CONTACT Dong Han, Email: donghan@sjtu.edu.cn , Department of Statistics, School of Mathematical Sciences, Shanghai Jiao Tong University, Shanghai, 200240, China.}  }
\affil{\textsuperscript{a,b}Department of Statistics, School of Mathematical Sciences, Shanghai Jiao Tong University, Shanghai, 200240, China.}
}

\maketitle

\begin{abstract}
In this paper, we propose an new the CUSUM sequential test (control chart, stopping time) with the observation-adjusted control limits (CUSUM-OAL) for monitoring quickly and adaptively the change in distribution of a sequential observations. We give the estimation of the in-control and the out-of-control average run lengths (ARLs) of the CUSUM-OAL test. The theoretical results are illustrated by numerical simulations in detecting $\alpha$ shifts of the extreme heavy-tailed distribution observations sequence.
\end{abstract}


\renewcommand{\thefootnote}{\fnsymbol{footnote}}
\footnotetext{${}^*$Supported by National Natural Science Foundation of China (11531001)}

\section{Introduction}

In order to quickly detect a change in  distribution of observations sequence without exceeding a certain false alarm rate,  a great variety of sequential tests have been proposed, developed and applied to various fields since \cite{Shewhart1931} proposed a control chart method, see, for example,\cite{SD1995}, \cite{BN1993}, \cite{Lai1995}, \cite{Lai2001}, \cite{Stoumbos2000}, \cite{CG2019}, \cite{Bersimis2007}, \cite{Montgomery2009}, \cite{Qiu2014}, \cite{Woodall2017}, \cite{Bersimis2018}.\\

One of popular used  sequential tests  is the following upper-sided CUSUM test which was proposed by \cite{Page 1954} .
\begin{equation}\label{equ1}
T_C(c)=\min \{n\geq 0:\,\, \max_{1\leq k\leq n}\sum_{i=n-k+1}^{n}Z_i \geq  c\},
\end{equation}

where $c>0$ is a constant control limit, $Z_i=\log [p_{v_1}(X_i)/p_{v_0}(X_i)]$, $p_{v_0}(x)$ and $p_{v_1}(x)$ are pre-change and post-change probability density functions respectively for a sequence of mutually independent observations $\{X_i,\, i\geq 1\}$, that is, there is a unknown change-point $\tau \geq 1$ such that $X_1, ..., X_{\tau-1}$ have the probability density function $p_{v_0}$, whereas, $X_{\tau}, X_{\tau+1}, ...$ have the probability density function $p_{v_1}$. By the renewal property of the CUSUM test $T_C$ we have $sup_{k\geq 1}\textbf{E}_k(T_C-k+1|T_C\geq k)=\textbf{E}_1(T_C)$ (see \cite{SD1995}, P.25), where $\textbf{E}_1(T_C)$ is the out-of-control average run length (ARL$_1$), $\textbf{P}_k$ and $\textbf{E}_k$ denote the probability and expectation respectively when the change from $p_{v_0}$ to $p_{v_1}$ occurs at the change-point $\tau=k$ for $k\geq 1$.\\

Though we know that the CUSUM test is optimal under Lorden's measure (see \cite{Moustakides1986} and \cite{Ritov1990}),  the out-of-control ARL$_1$ of the CUSUM test is not small, especially in detecting small mean shifts. In other words, the CUSUM test is insensitive in detecting  small mean shifts. Then, how to increase the sensitivity of the CUSUM test? Note that the control limit in the CUSUM test is a constant $c$ which does not depend on the observation samples.\\

Intuitively, if the control limit of the CUSUM test can become low as the samples mean of the observation sequence increases, then the alarm time of detecting the increasing mean shifts  will be greatly shortened. Based on this idea, by selecting a decreasing function $g(x)$  we  may define the ( upper-sided ) CUSUM chart  $T_C(cg)$ with  the observation-adjusted control limits $cg(\hat{Z}_n)$ ( abbreviated to the CUSUM-OAL chart ) in the following
\begin{equation}\label{equ2}
T_C(cg)=\min \{n\geq 0:\max_{0\leq k\leq n}\sum_{i=n-k+1}^{n}Z_i \geq cg(\hat{Z}_n)\},
\end{equation}
where $c>0$ is a constant and $\hat{Z}_n=\sum_{i=1}^nZ_i/n$.  In other words, the control limits $cg(\hat{Z}_n)$ of the CUSUM-OAL test  can be adjusted adaptively according to the observation information $\{\hat{Z}_n\}$.  Note that the control limits $cg(\hat{Z}_n)$  may be negative. In the special case, the CUSUM-OAL chart $T_C(cg)$ becomes into the conventional CUSUM chart $T_C(c)$ in (1) when $g\equiv 1$. Similarly, we can define a down-sided CUSUM-OAL test. In this paper, we consider only the upper-sided CUSUM-OAL test since the properties of  the down-sided CUSUM-OAL test can be obtained by the similar method.\\

In fact, there has been a lot of research work for improving the detection performance of the sequential tests by designing various time-varying (dynamic) control limits. \cite{Steiner 1999} and \cite{Celano and Castagliola 2018} showed that the EWMA scheme with dynamic control limits is more sensitive to  mean shifts of observations. By using the bootstrap approximations, \cite{Chatterjee and Qiu 2009} obtained a sequence of dynamic control limits that are determined by the conditional distribution of the CUSUM statistic. \cite{Lee2013} verified that  the individual sequential test with variable control limits  can quickly signal the mean change.\\

Another approach of using probability dynamic control limits for sequential tests was proposed by \cite{Margavio1995}. In recent years, many authors, for example, \cite{Verdier2017} ,\cite{Shen2013}, \cite{Zhang and Woodal 2015}, \cite{Zhang and Woodal 2017}, \cite{Huang2016}, \cite{Yang2017} and \cite{Sogandi2019} have used the probability dynamic control limits to improve the detection ability of the CUSUM and EWMA tests.\\

Heavy-tailed phenomena exist in many aspects of our lives such as physics, meteorology, computer science, biology, and finance. The theories about the heavy-tailed phenomenon have been used to study many aspects, for example, the magnitude of earthquakes, the diameter of lunar craters on the surface of the moon, the size of interplanetary fragments, and the frequency of words in human languages, and so on, see, for examples,  \cite{bib1}, \cite{Coo14}, \cite{bib3}, \cite{bib4}, \cite{tang2023}.We know that when  $\alpha \in (0, \, 1]$, the mean or variance is infinite, and when $\alpha \in (1, \, 2]$, the mean is finite but the variance is infinite.\\

Many of the traditional control chart methods are interesting in detecting on the changes of mean and variance. Because the mean or variance are infinite when the tail parameter $\alpha \in (0,1)$ of the extremely heavy-tailed distribution sequence, the mean or variance can not be used to construct the monitoring statistics. Considering that the extremely heavy-tailed parameter $\alpha $ centrally reacts to the distribution characteristics of the thick tail, and a small change in $\alpha$ will obviously change the degree of the heavy tail and the distribution characteristics, this paper adopts a change point method for the changes in its tail parameter $\alpha $ for the monitoring of the distribution sequence of the extremely heavy-tailed.\\

The monitoring of the extremely heavy-tailed parameters requires the control chart to be sensitive enough, which requires to construct very sensitive monitoring statistics. The cumulative sum method (CUSUM) is one of the most widely used control chart monitoring methods, but this method is not sensitive to small displacement variations.\\

However, unlike all the dynamic non-random control limits mentioned above, our proposed the dynamic random control limits or the observation-adjusted control limits depend on the real-time observations, so it can improve the detection ability of the CUSUM test adaptively. The main purpose of the present paper is to show the good detection performance of the CUSUM-OAL test and to give the estimation of its the in-control and out-of-control ARLs.\\

The paper is organized as follows. In Section 2, we first define two sequences of the CUSUM-OAL tests and prove that one of the two sequences of CUSUM-OAL tests converges to the optimal test, another sequences of CUSUM-OAL tests converges to a combination of the optimal test and the CUSUM test.  The estimation of the in-control and out-of-control ARLs of the CUSUM-OAL tests and their comparison are given in Section 3. The detection performances of the three CUSUM-OAL tests and the conventional CUSUM test are illustrated in Section 4 by comparing their numerical out-of-control ARLs. Section 5 provides some concluding remarks. Proofs of the theorems are given in the Appendix.\\

\section{ CUSUM-OAL test and estimation and comparison of its ARL.}
For comparison, the in-control $\textbf{ARL}_{0}$ of all candidate sequential tests are constrained to be equal to the same desired level of type I error,  the test with the lowest out-of-control $\textbf{ARL}_{v}$  has the  highest power or the fastest monitoring (detection) speed.\\

In this section we will give  an estimation  of the  ARLs of the following  CUSUM-OAL test that can be written as
\begin{equation}
T_C(cg)=\min \{n\geq 1:\max_{1\leq k\leq n}\sum_{i=n-k+1}^{n}Z_i
\geq cg(\hat{Z}_n(ac))\}, \label{f07}
\end{equation}
where $g(.)$ is a decreasing function, $\hat{Z}_n(ac)=\frac{1}{j}\sum_{i=n-j+1}^nZ_i, j=\min\{n, [ac+1]\}$ for $a>0, c>0$,  and $[x]$ denotes the smallest integer greater than  or equal to $x$. Here $\hat{Z}_n(ac)$ is a sliding average of the statistics, $Z_i, n-j+1\leq i\leq n$,  which will become $\hat{Z}_n=\frac{1}{n}\sum_{i=1}^nZ_i$ when $a=\infty$.\\

Next we discuss on the the post-change probability distribution in order to estimate the ARLs of  $T_C(cg)$.\\

Usually we rarely know the post-change probability distribution $P_{v}$ of the observation process before it is detected. But the possible change domain   and its boundary (including the size and form of the boundary) about $v$  may be determined by engineering knowledge, practical experience or statistical data. So we may assume that the region of parameter space $V$ and a probability distribution $Q$ on $V$ are known.  If we have no prior knowledge of the possible value of $v$ after the change time $\tau$,  we may assume that $v$ occurs equally on $V$,  that is, the probability distribution $Q$ is an  equal probability distribution (or uniform distribution ) on $V$.  For example, let $P_{v}$ be the normal distribution and $v=(\mu, \sigma)$, where $\mu$ and $\sigma $ denote the mean and standard deviation respectively, we can take the set $V=\{(\mu, \sigma): \mu_1 \leq \mu \leq \mu_2, 0< \sigma_1\leq \sigma \leq \sigma_2\}$ and $Q$ is subject to the uniform distribution $U(V)$ on $V$ if $v$ occurs equally on $V$, where the numbers $\mu_1, \mu_2, \sigma_1$ and $\sigma_2$ are known. It means that we know the domain of the possible post-change distributions, $P_v, v\in V$, i.e., the boundary $\partial{V}$ of the parameter space $V$ is known.\\

Let the change-point $\tau =1$. As in \cite{Lai1998}, we consider the mixture likelihood ratio statistics
\begin{eqnarray*}
Z_k=\log[\frac{P_{v_1}(X_k)}{P_{v_0}(X_k)}], \,\, k\geq 1,
\end{eqnarray*}
with
\begin{eqnarray*}
P_{v_1}(X_k)=\int_{V}P_{v}(X_k)Q(dv).
\end{eqnarray*}
Here, $P_{v_0}(x)=P_{v_0}(X_k=x)$ and $P_v(x)=P_{v}(X_k=x)$ if all $X_k, k\geq 1,$ are discrete random variables.  When $X_k, k\geq 1,$ are continuous random variables, $P_{v_0}$ and $P_v$ will be replaced by their density functions $p_{v_0}$ and $p_{v}$ respectively. Note that $P_{v_1}$ can be considered as a known reference post-change probability
distribution.  In order to guarantee the validity of $Z_k$, we assume that
\begin{eqnarray*}
P_{v_0}(x)>0 \,\, \text{ if and only if } \,\, P_{v_1}(x)>0.
\end{eqnarray*}
Define $P_{v_1}(x)/P_{v_0}(x)=1$ when $P_{v_1}(x)=P_{v_0}(x)=0$.\\

Next we shall divide the parameter space $V$ into three subsets $V^{+}$, $V^0$ and $V^{-}$ by the  Kullback-Leibler information
distance.  Let
\begin{eqnarray*}
V^{-}=\{v: E_{v}(Z_1)<0\},\,\,\,\,\,\,V^{0}=\{v:
E_{v}(Z_1)=0\},\,\,\,\,\,\,V^{+}=\{v: E_{v}(Z_1)>0\}
\end{eqnarray*}
where $E_{v}(Z_1)=I(P_{v}|P_{v_0})-I(P_{v}|P_{v_1})$ and
\begin{eqnarray*}
I(P_{v}|P_{v_0})=E_{v}(\log[\frac{P_{v}(X_1)}{P_{v_0}(X_1)}]),\,\,\,\,\,\,\,\,
I(P_{v}|P_{v_1})=E_{v}(\log[\frac{P_{v}(X_1)}{P_{v_1}(X_1)}])
\end{eqnarray*}
are two Kullblak-Leibler information distances  between $P_{v}$, $P_{v_0}$ and $P_{v}$, $P_{v_1}$. Since $I(p|q)= 0$ if and only if $p=q$, where $p$ and $q$ are two probability measures, it follows that $E_{v_0}(Z_1)=-I(P_{v_0}|P_{v_1})$, and therefore, $v_0 \in V^{-}$ when $P_{v_1}\neq P_{v_0}$. When $v \in V^{-}$, i.e., $I(P_{v}|P_{v_0})<I(P_{v}|P_{v_1})$, it means that $P_{v}$ is closer to $P_{v_0}$ than to $P_{v_1}$ according to the Kullblak-Leibler information distance. There is a similar explanation for $v \in V^{+}$ or $\in V^{0}$.\\

Suppose the post-change distribution $P_{v}$ and the function $g(x)$ satisfy the following conditions:\\
\textbf{(I)} The probability $P_{v}$ is not a point mass at $E_{v}(Z_1)$ and $P_v(Z_1>0)>0$.\\
\textbf{(II)} The  moment-generating function
$h_{v}(\theta)=E_{v}(e^{\theta Z_1})$ satisfies
$h_{v}(\theta)<\infty $ for some $\theta >0$. \\
\textbf{(III)} The function  $g(x)$ is decreasing, its second order derivative function $g''(x)$ is continuous and bounded, and there is a positive number $x^*$ such that $g(x^*)=0$.\\

Let $\tilde{Z}_1=Z_1+|g'(\mu)|(Z_1-\mu)/a$, $\tilde{h}(\theta)=\textbf{E}_v(e^{\theta \tilde{Z}_1})$ and 
\begin{eqnarray*}
H_v(\theta)=\frac{au}{g(\mu)}\ln \tilde{h}_v(\theta) +
(1-\frac{au}{g(\mu)})\ln h_{v}(\theta)
\end{eqnarray*}
for $\theta \geq 0$, where $E_{v}(Z_1)=\mu <0$, $a\leq g(\mu)/u$ and $u=H'_v(\theta^*_v)$, where $\theta^*_v >0$ satisfies
$H_v(\theta^*_v)=0$.  Note that $H_v(\theta)$ is a convex function
and $H'_v(0)=\mu<0$. It follows that there is a unique positive
number $\theta^*_v >0$ such that $H_v(\theta^*_v)=0$.\\

The following theorem gives the estimations of the $\textbf{ARL}_{v}$ of $T_c(g)$ for a large $c$.
\par
\bigskip
\begin{theorem}\label{theorem3}
Suppose that the conditions (I)-(III) hold. Let $P_{v_1} \neq
P_{v_0}$, $E_{v_0}(Z_1)=\mu_0 <0$ and $a\leq g(\mu)/u$ for $\mu_0\leq \mu <0$.  Then, for a large  control limit $c$,  we have
\begin{eqnarray}
\small{\textbf{ARL}_{v}(T_c(g))= \begin{cases}
 A\frac{e^{c\theta_{v}^*g(\mu)}}{|\mu|}(1+o(1))&\text{if $E_v(Z_1)<0$, or\, $v\in V^{-}$}\\
 B\frac{(cg(0))^2}{\sigma^2}(1+o(1))&\text{if $E_v(Z_1)=0$, or\, $v\in V^{0}$}\\
     \frac{cg(\mu)}{\mu}(1+o(1))&\text{if $E_v(Z_1)>0$, or\, $v\in V^{+}$}
 \end{cases} }\label{f11}
\end{eqnarray}
where $\sigma^2=E_v(Z_1^2)$ when $E_v(Z_1)=0$, the two numbers $A$ and $B$ satisfy $1/bc\leq A\leq cg(\mu)/u$ and $(8\ln c )^{-1}\leq
B\leq (1-\Phi(1))^{-1}$, respectively, $u=H'_v(\theta^*_v)$  and $b$
is a positive number defined by
\begin{eqnarray}
b=\inf \{x>g(\mu)/u:
\theta(\frac{1}{x})-xH_{v}(\theta(\frac{1}{x}))\geq 2\theta^*_v\}
\label{f08}
\end{eqnarray}
where $\theta(1/x)$ satisfies $1/x=H'_v(\theta(1/x))$ and $\Phi(x)$
is the distribution function of the standard normal distribution.
\end{theorem}

Note that the following function
\begin{eqnarray*}
\Theta(x)=\theta(\frac{1}{x})-xH_{v}(\theta(\frac{1}{x}))-2\theta^*_v
\end{eqnarray*}
satisfies that $\Theta(1/u)=\theta(u)-2\theta^*_v=-\theta^*_v$,
where $\theta(u)=\theta^*_v$,
\begin{eqnarray*}
\Theta'(x)=-\frac{1}{x^2}\theta'(\frac{1}{x})-H_{v}(\theta(\frac{1}{x}))+\frac{1}{x}\theta'(\frac{1}{x})H_v'(\theta(\frac{1}{x}))=-H_{v}(\theta(\frac{1}{x}))
\end{eqnarray*}
and therefore,  $\Theta'(\theta(u))=-H(\theta(u))=-H(\theta^*_v)=0$, $\Theta'(\theta(1/x))>0$ for $x>1/u$ and $\Theta'(\theta(1/x))<0$ for $x>1/u$. Hence, there exists a positive number $b$ defined in(\ref{f08}).\\

It can be seen, the main part of $\textbf{ARL}_{v}(T_{c}(g))$ will be an exponential function, square function, and linear function of $c$ when  the process $\{Z_k: k\geq 0\}$ has  no change or a "small
change", a "medium change"  and  a "large change" from $P_{v_0}$ to $P_{v}$, respectively.  Here, the "small change" ($v\in V^-$) means that $P_{v}$ is closer to $P_{v_0}$ than to $P_{v_1}$, i.e., $I(P_{v}|P_{v_0})<I(P_{v}|P_{v_1})$, and the "large change" is just the opposite.  The "medium change" ($v\in V^0$) corresponds to $I(P_{v}|P_{v_0})=I(P_{v}|P_{v_1})$.

\bigskip
\begin{remark}\label{remark3}
Let $g\equiv 1$. By using Wald's identity and the martingale method, we can obtain the following well known ARLs estimation of the CUSUM test $T_C$ (see \cite{SD1995} and \cite{Poor2008})
\begin{eqnarray}
\textbf{ARL}_{v}(T_{C}(c))= \begin{cases}
 \frac{e^{cs_v^*}}{|\mu|}(1+o(1)) &\text{ if\ $E_v(Z_1)=\mu<0$}  \\
 \frac{c^2}{\sigma^2}(1+o(1))                  &\text{ if \  $E_v(Z_1)=\mu=0$} \\
     \frac{c}{\mu}(1+o(1)) &\text{ if \  $E_v(Z_1)=\mu>0$}
 \end{cases} \label{f09}
\end{eqnarray}
where $s^*_{v}$ satisfies $h_v(s^*_{v})=1$ and $s^*_{v_0}=1$.\\
\end{remark}

Note that $\theta^*_{v}=s^*_v$ when $g\equiv 1$ since $H_v(\theta)=\ln h_v(\theta)$  when $g\equiv 1$.\\

In this paper, we will use another  method to prove Theorem \ref{theorem3} since Wald's identity and the martingale method do not hold or can not work for showing the  ARLs estimation of the test $T_c(g)$ when $g$ is not constant.\\

Next we compare the detection performance of the CUSUM-OAL  test ($\textbf{ARL}_{v}(T_{c'}(g))$)  with that of  the CUSUM test ($\textbf{ARL}_{v}(T_{C}(c))$) by using (\ref{f11}).\\

Let $\textbf{ARL}_{v_0}(T_{c'}(g))=\textbf{ARL}_{v_0}(T_{C}(c))$ forlarge $c'$ and $c$. We have $c=c'\theta_{v_0}^*g(\mu_0) +o(1)$.
Hence
\begin{eqnarray*}
\textbf{ARL}_{v}(T_{C}(c))>\textbf{ARL}_{v}(T_{c}(g))
\end{eqnarray*}
for $s^*_{v}\theta^*_{v_0}> g(\mu)\theta^*_v/g(\mu_0)$ when $\mu_0<\mu <0$ and for $\theta^*_{v_0}>g(\mu)/g(\mu_0)$ when $\mu\geq 0$.  This means that $\textbf{ARL}_{v}(T_{c}(g))$ can be smaller than $\textbf{ARL}_{v}(T_{C}(c))$ as long as $g(\mu)/g(\mu_0)$ is small for all $\mu >\mu_0$.\\

Furthermore we have the following corollary.\\

\begin{corollary}\label{corollary}
Let $g_1$ and $g_2$ satisfy the condition (III). Suppose that $\textbf{ARL}_{v_0}(T_{c_1}(g_1))\sim \textbf{ARL}_{v_0}(T_{c_2}(g_2))$ for large control limits $c_1$ and $c_2$, that is, $c_1\theta_1^*(v_0)g_1(\mu_0)=c_2\theta_2^*(v_o)g_2(\mu)+o(1)$. Then
\begin{eqnarray*}
\textbf{ARL}_{v}(T_{c_1}(g_1))< \textbf{ARL}_{v}(T_{c_2}(g_2))
\end{eqnarray*}
if and only if
\begin{eqnarray*}
\frac{\theta_1^*(v)g_1(\mu)}{\theta_1^*(v_0)g_1(\mu_0)}< \frac{\theta_2^*(v)g_2(\mu)}{\theta_2^*(v_0)g_2(\mu_0)}
\end{eqnarray*}
for $\mu_0\leq \mu<0$, and
\begin{eqnarray*}
\frac{g_1(\mu)}{\theta_1^*(v_0)g_1(\mu_0)}< \frac{g_2(\mu)}{\theta_2^*(v_0)g_2(\mu_0)}
\end{eqnarray*}
for $\mu\geq 0$.
\end{corollary}

\begin{example}\label{example2}
Let $X_0 \sim N(0, 1)$ and, after the change point $\tau=1$,  $X_k \sim N(1, 1)$ for $k\geq \tau=1$. It follows that $Z_k=X_k-1/2$, $\mu_0=-1/2$ and $\textbf{E}_v(Z_1)=\mu=(v-1/2)$. In this case, the positive numbers $\theta_{v}^*$ and $u$ in Theorem \ref{theorem3} can be written by
\begin{eqnarray*}
\theta_{v}^*=\frac{2a|\mu| g(\mu)}{ag(\mu)+|\mu g'(\mu)|(2a+|g'(\mu)|)}
\end{eqnarray*}
and $u=|\mu|$. When $g'(\mu)\equiv 0$ or $g$ is constant, we have $\theta_{v}^*=2|\mu|=1-2v$, which is qual to $s^*_v$ in (\ref{f09}). Taking large $c$ and $c'$ such that $c=c'g(\mu_0)\theta^*_{v_0}$,
then, by Corollary \ref{corollary}, we have
\begin{eqnarray*}
\textbf{ARL}_{v}(T_{C}(c))>\textbf{ARL}_{v}(T_{c'}(g))
\end{eqnarray*}
for $\textbf{E}_v(Z_1)=\mu \geq 0$, as long as
\begin{eqnarray}
\frac{a g^2(\mu_0)}{ag(\mu_0)+|\mu_0 g'(\mu_0)|(2a+|g'(\mu_0)|)}\,>\,g(\mu). \label{f10}
\end{eqnarray}
\end{example}

\bigskip

\begin{remark}
It was shown by \cite{Moustakides1986} and \cite{Ritov1990} that the performance of the CUSUM test $T_C$ in detecting the mean shift with the reference value $\delta$ is optimal in  Lorden's measure if the real mean shift is $\delta$. \cite{Han and Fugee 2004} and \cite{Han and Fugee 2006} showed that the ARL of the CUSUM test with the reference value $\delta$ is the smallest among all commonly used tests such as GLR, Cuscore, RFCuscore, EWMA,  GEWMA and the optimal EWMA if the real mean shift is $\delta$ when $\tau =1$. However, example \ref{example2} shows that as long as $g(\mu_1)$ satisfies (\ref{f10}), i.e., $\textbf{E}_{v_1}(Z_1)=\mu_1$, the ARL of the CUSUM-OAL test $T_{c}(g)$ is smaller than that of the CUSUM test $T_C$ with the reference value $\delta =1$ in detecting mean shift $v=v_1=\delta=1$ for a large $c$ when $\tau =1$.
\end{remark}

It can be seen that $\textbf{ARL}_{v}(T_c(g))\to \infty$ as $c\to \infty$ for $g(\mu)>0$. When $g(\mu)<0$, the control limit $cg(\hat{Z}_n)$ in $T_g(c)$ may be negative. The following theorem shows that $\textbf{ARL}_{v}(T_c(g))$ is bounded for any $c>0$ when $g(\mu)<0$.\\

\bigskip

\begin{theorem}\label{theorem4}
Let $\hat{Z}_n(ac)=\hat{Z}_n=n^{-1}\sum_{i=1}^nZ_i$ and $\textbf{ARL}_{vk}(T_c(g)) =\textbf{E}_{vk}(T_c(g)-k+1)^+$ for $k \geq 1$. Also let $\textbf{E}_v(Z_1)=\mu>0$ and $g(\mu)<0$. Then, for any $c>0$,
\begin{eqnarray*}
\textbf{ARL}_{v}(T_c(g)) \leq B
\end{eqnarray*}
for $k=1$ and
\begin{eqnarray*}
\textbf{ARL}_{vk}(T_c(g)) \leq (a_0+1)(k-1)\textbf{P}_0(T_c(g)\geq k)+\frac{2e^{-(a_0+1)(k-1)b}}{1-e^{-b}}
\end{eqnarray*}
for $k>1$, where
\begin{eqnarray*}
B&=&\frac{e^{2b}+1}{e^{2b}-1},\,\,\,\,\,\,\,\,\,\,\,\,\,\,\,\,\,\,\,\,\,\,\,\,\,\,\,\,\,a_0=\min\{\frac{b}{d},\,\,\,d\}\\ d&=&|\mu_0-\mu^*-\ln M_0(\theta^*)|,\,\,\,\,\,\,\,\,\,\,\,\,\,\, b=\frac{\theta^* (\mu-\mu^*)-\ln M_v(\theta^*)}{2},
\end{eqnarray*}
and $M_v(\theta)=E_v(e^{\theta(\mu-Z_1)})$,  $\mu^*=\max\{a^*,\, 0\}<\mu$,  $a^*$ satisfies $g(a^*)=0$,  and  $\theta^*>0$ satisfies
\begin{eqnarray*}
\frac{M'_v(\theta^*)}{M_v(\theta^*)}=\mu-\mu^*.
\end{eqnarray*}
\end{theorem}

Theorem \ref{theorem4} shows that $\textbf{ARL}_{vk}(T_c(g)) $ is bounded for any $c>0$ when $g(\mu)<0$. But the commonly used detection tests, such as, the CUSUM, EWMA, GLRT, CUSCORE, Shiryaev-Roberts tests,
etc.,  satisfy  $\textbf{ARL}_{vk}\to \infty$ as $\textbf{ARL}_{0} \to \infty$, since all their control limits are positive constants and go to $\infty$  when $\textbf{ARL}_{0}\to \infty$.\\

\begin{remark}\label{remark5}
The four nonnegative constants, $B, a_0, d$ and $b$ in Theorem \ref{theorem4} do not depend on $c$.
\end{remark}

\bigskip

\section{NUMERICAL SIMULATION}
Let  $X_k, k\geq 1,$ be i.i.d.  with extremely heavy-tailed distribution function $F(x)=1-1/x^{\alpha}$ and probability density function $f(x)=\alpha/x^{1+\alpha}$ for $x\geq 1$, $0<\alpha<1$. There is a unknown change-point $\tau \geq 1$ such that $X_1, ..., X_{\tau-1}$ have the probability density function $f_{\alpha_0}$, whereas, $X_{\tau}, X_{\tau+1}, ...$ have the probability density function $f_{\alpha_1}$. We consider the mixture likelihood ratio statistics
\begin{eqnarray*}
Z_k=\log\frac{f_{\alpha_1}}{f_{\alpha_0}}=\log{\frac{\alpha_1}{\alpha_0}}+(\alpha_0 -\alpha_1 )\log X_k,
\end{eqnarray*}\\
where $X_k \geq 1$, $0<\alpha_0<1$, $0<\alpha_1<1$.\\ 

We know that
\begin{eqnarray*}
EZ_k=\log{\frac{\alpha_1}{\alpha_0}}+(\alpha_0 -\alpha_1 )E\log X_k.
\end{eqnarray*}\\

Next, in order to speed up the monitoring speed of CUSUM test, we present the next CUSUM-OAL test when $\tau=1$. Taking the serie of decreasing functions $g_{u}(x)=1-u(x-\mu_0)$ for $ x > \mu_0$ and ${g_{u}}(x)=1$ for  $x \leq \mu_0$. Where $-\infty < x < \infty$, $u\geq 0$, we define the CUSUM-OAL test in the following
\begin{eqnarray}
T_C(cg_{u})=\min \{n\geq 0:\max_{1\leq k\leq n}\sum_{i=n-k+1}^{n}Z_i\geq  cg_{u}(\hat{Z}_n)\},
\end{eqnarray}\\
where $c>0$ is a constant, $g_{u}(\hat{Z}_n)=1-u(\hat{Z}_n-\mu_0)$, $\hat{Z}_n=n^{-1}\sum_{k=1}^nZ_k$, $EZ_k=\mu_0$ .\\

\bigskip
\par
\par
\footnotesize \setlength{\tabcolsep}{6pt}
\begin{table}
  \centering
\scalebox{0.8}{
\begin{tabular}{|c|l|l|l|l|l|l|}
\multicolumn{5}{l}{\normalsize \textbf{Table 1}.\,\,\,\,\,\ \textbf{ARLs} of $T_C(c)$ and $T_C(cg_{u})$ when $\tau=1$,$\alpha_0=0.90$.}\\[5pt]
\hline
&\multicolumn{3}{c|}{\textbf{$ARL_0=300$}}&\multicolumn{3}{c|}{\textbf{$ARL_0=500$}}\\\cline{2-7}
              & $T_C(c)$ & \multicolumn{2}{c|}{$T_C(cg_{u})$} & $T_C(c)$ & \multicolumn{2}{c|}{$T_C(cg_{u})$} \\\cline{2-7}
$\alpha_1$    &                & $u=0.50$       &  $u=5$           &           &  $u=0.50$  & $u=5$ \\
              & $c=10.89$        & $c=11.32$       & $c=12.49$        & $c=12.48$ &  $c=12.92$ & $c=14.09$\\\hline
$0.90$        & $300.89(291.33)$ & $300.37(335.71)$ & $300.77(462.17)$ & $500.62(487.90)$&500.73(560.33)&500.45(502.68) \\\hline
$0.70$        & $51.12(44.23)$ & $41.99(47.67)$ & $29.75(49.31)$ &$65.40(56.89)$&53.05(60.71)&36.49(61.97) \\\hline
$0.50$        & $15.28(10.65)$ & $9.88(10.99)$ & $5.75(9.41)$ &$17.64(11.93)$&10.99(12.52)&5.97(10.06) \\\hline
$0.30$        & $6.16(3.63)$ & $3.20(3.21)$ & $2.07(2.29)$ &$6.85(3.88)$&3.34(3.47)&2.10(2.38) \\\hline
\end{tabular}\label{table1}}
\end{table}

\normalsize

Numerical simulations above show that, as can be seen from {\textbf{Table $1$}}, whether {\textbf{$ARL_0=300$}} or {\textbf{$ARL_0=500$}}, $T_C(c)$'s test result is much better than $T_C(cg_u)$'s, the out-control ARLs of the CUSUM-OAL tests are significantly smaller than the out-control ARLs of the CUSUM test. But, as a whole, the CUSUM-OAL tests have bigger standard deviations than the CUSUM test in the in-control state. Both theoretical estimations and numerical simulations show that the CUSUM-OAL tests perform much better than the CUSUM test when $ \tau=1$.

\section{ Conclusion}
The contributions of this paper can be summarized to the following two aspects.\\

(1) To enhance the sensitivity of the CUSUM test for detecting the distribution change, we present CUSUM-type test with a real-time observation control limit (CUSUM-OAL). Numerical simulations show that CUSUM-OAL's test result is much better than CUSUM's, the out-control ARLs of the CUSUM-OAL tests are significantly smaller than the out-control ARLs of the CUSUM tests.\\

(2) The extreme heavy-tailed distribution mainly depends on parameter $\alpha$. By CUSUM-OAL test, we designed a better detection method, which can be more sensitive to detect small changes in the parameter of the heavy-tailed distribution, which plays an important role in analyzing the change point monitoring of the heavy-tailed distribution, The extreme thick tail distribution mainly depends on parameter. We designed a better detection method, which can be more sensitive to detect small changes in the parameter of the thick tail distribution, which plays an important role in analyzing the change point monitoring of the thick tail distribution, this is proved by numerical simulation.

\par
\newpage
\vskip 3cm
\textbf{APPENDIX : Proofs of Theorems }
\bigskip

\textbf{Proof of Theorem \ref{theorem3}.}  Let $v\in V^{-}$. We first prove that
\begin{equation}
e^{cg(\mu)\theta^*_v(1+o(1))}/bc\leq E_{v}(T_c(g))\leq
cu^{-1}g(\mu)e^{cg(\mu)\theta^*_v(1+o(1))}
\end{equation}
for a large $c$.

Next we first prove the upward inequality of (36). Let $m_1=\left\lceil cu^{-1}g(\mu) \right\rceil$, $m_k=\left\lceil km_1\right\rceil $ for $k\geq 0$ and  $m= \left\lceil tm_1\exp\{cg(\mu)\theta^*_v(1+o(1))\}\right\rceil $ for $t>0$, where $ \left\lceil x\right\rceil $ denotes the smallest integer
greater than or equal to $x$. Without loss of generality, the number $ \left\lceil x\right\rceil $ will be replaced by $x$ in the following when $x$ is large.   It follows that
\begin{eqnarray}
P_{v}(T_c(g)>m)&=&P_{v}(\sum_{i=n-k+1}^{n}Z_{i}<cg(\hat{Z}_n(ac)), \
\ \ 1\leq k\leq n, 1 \leq
n\leq m) \nonumber\\
&\leq & P_{v}(\sum_{i=m_{j}-m_1+1}^{m_j}Z_{i}<cg(\hat{Z}_{m_j}(ac)), \,\, 1\leq j\leq m/m_1 )\nonumber\\
&= & [P_v(\sum_{i=1}^{m_1}Z_{i}<cg(\hat{Z}_{m_1}(ac))]^{m/m_1}
\end{eqnarray}
for a large $c$,  where
$\hat{Z}_{m_j}(ac))=(ac)^{-1}\sum_{i=m_j-ac+1}^{m_j}Z_i$ and the
last quality holds since the events
\begin{eqnarray*}
\{\sum_{i=m_{j}-m_1+1}^{m_j}Z_{i}<cg(\hat{Z}_{m_j}(ac))\},
\end{eqnarray*}
$1\leq j\leq m/m_1$, are mutually independent and have an identity distribution.  Since $\hat{Z}_{m_1}(ac)-\mu \to 0 (a.s.)$ and
$ac(\hat{Z}_{m_1}(ac)-\mu )^2 \Rightarrow \chi^2$ ($\chi^2$ -distribution)  as $c\to \infty$, it follows that
\begin{eqnarray*}
g(\hat{Z}_{m_1}(ac))=g(\mu)+g'(\mu)(\hat{Z}_{m_1}(ac)-\mu)+O(1/c)
\end{eqnarray*}
and
\begin{eqnarray*}
P_v(\sum_{i=1}^{m_1}Z_{i}<cg(\hat{Z}_{m_1}(ac))=P_v(\sum_{i=m_1-ac+1}^{m_1}\tilde{Z}_{i}+\sum_{i=1}^{m_1-ac}Z_{i}<c(g(\mu)+O(1/c)))
\end{eqnarray*}
for a large $c$, where $\tilde{Z}_{i}=Z_i+a^{-1}|g'(\mu)|(Z_i-\mu)$.
Let
\begin{eqnarray*}
\tilde{h}_v(\theta)=E_v(e^{\theta
\tilde{Z}_1}),\,\,\,h_{v}(\theta)=E_{v}(e^{\theta Z_1}).
\end{eqnarray*}
and
\begin{eqnarray*}
H_v(\theta)=\frac{au}{g(\mu)}\ln \tilde{h}_v(\theta) +
(1-\frac{au}{g(\mu)})\ln h_{v}(\theta).
\end{eqnarray*}
Note that $H_v(\theta)$ is a convex function and $H'_v(0)=\mu<0$.
This means that there is a unique positive number $\theta^*_v >0$ such that $H_v(\theta^*_v)=0$. Let $u=H'_v(\theta^*_v).$ It follows
from (A.9) that
\begin{eqnarray*}
&&P_v(\sum_{i=m_1-ac+1}^{m_1}\tilde{Z}_{i}+\sum_{i=1}^{m_1-ac}Z_{i}\geq c(g(\mu)+O(1/c)))\\
&&=P_v(\sum_{i=m_1-ac+1}^{m_1}\tilde{Z}_{i}+\sum_{i=1}^{m_1-ac}Z_{i}\geq m_1u(1+O(1/c)))\\
&&\geq \exp\{-m_1(\theta u' -H_v(\theta)+\frac{1}{m_1}\log (F^{m_1}_{\theta}(m_1u')-F^{m_1}_{\theta}(m_1u))+o(1))\}\\
&&=\exp\{-cg(\mu)(\theta\frac{u'}{u} -\frac{1}{u}H_v(\theta)+o(1))\}
\end{eqnarray*}
for a large $c$. Taking $\theta \searrow \theta^*_v$ and $u'\searrow u$, we have
\begin{eqnarray*}
P_v(\sum_{i=m_1-ac+1}^{m_1}\tilde{Z}_{i}+\sum_{i=1}^{m_1-ac}Z_{i}<
c(g(\mu)+O(1/c)))\leq 1-\exp\{-cg(\mu)\theta^*_v(1+o(1))\}
\end{eqnarray*}
for a large $c$. Thus, by (A.11) we have
\begin{eqnarray}
P_{v}(T_c(g)>m)\leq [ 1-\exp\{-cg(\mu)\theta^*_v(1+o(1))\}]^{m/m_1}
\to e^{-t}.
\end{eqnarray}
as $c\to \infty$. By the properties of exponential distribution, we
have
\begin{eqnarray*}
E_v(T_c(g)))\leq cu^{-1}g(\mu)e^{cg(\mu)\theta^*_v(1+o(1))}
\end{eqnarray*}
for a large $c$.

To prove the downward inequality of (A.10), let
\begin{eqnarray*}
U_m &=&\{\sum_{i=n-k+1}^{n} Z_{i}<cg(\hat{Z}_n(ac)),\;\; 1\leq k\leq  ac-1, \, bc\leq n \leq m \}\\
V_m &=&\{\sum_{i=n-k+1}^{n} Z_{i}<cg(\hat{Z}_n(ac)),\;\; ac\leq k\leq  bc-1, \, bc\leq n \leq m \}\\
W_m &=&\{\sum_{i=n-k+1}^{n} Z_{i}<cg(\hat{Z}_n(ac)), \;\; bc\leq k\leq n, \; bc\leq n\leq m \}\\
S_{bc}&=&\{\sum_{i=n-k+1}^{n} Z_{i}<cg(\hat{Z}_n(ac)), \;\; 1\leq
k\leq n, \; 1\leq n\leq bc-1 \},
\end{eqnarray*}
where $b$ is defined in (\ref{f08}) and without loss of generality, we assume that $b>a$. Obviously, $\{T_c(g)>m\}=U_m V_m W_{m}S_{bc}$.

Let $k=xcg(\mu)$. By Chebyshev's inequality, we have
\begin{eqnarray*}
P_v(\sum_{i=n-k+1}^{n} Z_{i}<cg(\hat{Z}_n(ac)))&=&P_v\Big(\sum_{i=n-k+1}^{n} \tilde{Z}_{i}+\sum_{i=n-ac+1}^{n-k}\tilde{\tilde{Z}}_i< cg(\mu)(1+o(1))\Big)\\
&\geq &1-\exp\{-cg(\mu)(\theta -x\tilde{H}_v(\theta)+o(1))\}
\end{eqnarray*}
for $1\leq k\leq  ac-1, \, bc\leq n \leq m$, where
$\tilde{\tilde{Z}}_i=-g'(\mu)(Z_{i}-\mu)/a$ and
\begin{eqnarray*}
\tilde{H}_v(\theta)=\ln \tilde{h}_v(\theta) + (\frac{ac}{k}-1)\ln
\hat{h}_{v}(\theta),\,\,\,\,\, \hat{h}_{v}(\theta)=E_v(e^{\theta
\tilde{\tilde{Z}}_i}).
\end{eqnarray*}
Since $\tilde{H}_v(\theta)$ and  $H_v(\theta)$ are two convex
functions and
\begin{eqnarray*}
&&\tilde{H}'_v(0)-H'_v(0)=0,\\
&&\tilde{H}''_v(0)-H''_v(0)=\sigma^2[(1+\frac{g'(\mu)}{a})^2+(\frac{ac}{k}-1)-\frac{au}{g(\mu)}(1-\frac{g'(\mu)}{a})^2+\frac{au}{g(\mu)}-1]>0,
\end{eqnarray*}
it follows that $\tilde{\theta}_v^* \geq \theta_v^*$, where
$\tilde{\theta}_v^*$ and $\theta_v^*$ satisfy
$\tilde{H}_v(\tilde{\theta}_v^*)=H_v(\theta_v^*)=0$. Hence
\begin{align}
P_v(\sum_{i=n-k+1}^{n}
\tilde{Z}_{i}+\sum_{i=n-ac+1}^{n-k}\tilde{\tilde{Z}}_i&<cg(\mu)(1+o(1)))\nonumber\\
&\geq 1-\exp\{-cg(\mu)\theta_v^*(1+o(1))\}
\end{align}
for $1\leq k\leq  ac-1, \, bc\leq n \leq m$. Similarly, we can get
\begin{align}
P_v(\sum_{i=n-ac+1}^{n} \tilde{Z}_{i}+\sum_{i=n-k+1}^{n-ac}Z_{i}&<
cg(\mu)(1+o(1)))\nonumber\\
&\geq 1-\exp\{-cg(\mu)\theta_v^*(1+o(1))\}
\end{align}
for $ac\leq k\leq  bc-1, \, bc\leq n \leq m$, and
\begin{eqnarray}
P_v(\sum_{i=n-ac+1}^{n} \tilde{Z}_{i}+\sum_{i=n-k+1}^{n-ac}Z_{i}&<&
cg(\mu)(1+o(1)))\nonumber\\
&\geq& 1-\exp\{-2cg(\mu)\theta_v^*(1+o(1))\}
\end{eqnarray}
for $bc\leq k\leq  n, \, bc\leq n \leq m$.

Let $m=tcg(\mu)\theta_v^*/bc$ for $t>0$. By (A.13), (A.14), (A.15)
and Theorem 5.1 in Esary, Proschan and Walkup (1967) we have
\begin{eqnarray*}
P_v(U_mV_m)&\geq & \prod_{n=bc}^{m}\prod_{k=1}^{ bc-1}P_v(\sum_{i=n-k+1}^{n} \tilde{Z}_{i}-\frac{g'(\mu)}{a}\sum_{i=n-ac+1}^{n-k}(Z_{i}-\mu)< cg(\mu)(1+o(1)))\\
&\geq& [1-\exp\{-cg(\mu)\theta_v^*(1+o(1))\}]^{bc m} \to e^{-t}
\end{eqnarray*}
and
\begin{eqnarray*}
P_v(W_m)&\geq & \prod_{n=bc}^{m}\prod_{k=bc}^{ n}P_v(\sum_{i=n-ac+1}^{n} \tilde{Z}_{i}+\sum_{i=n-k+1}^{n-ac}Z_{i}<cg(\mu)(1+o(1)))\\
&\geq& [1-\exp\{-2cg(\mu)\theta_v^*(1+o(1))\}]^{(m-bc)^2} \to 1
\end{eqnarray*}
as $c\to +\infty$.

Finally,
\begin{eqnarray*}
P_v(S_{bc})&\geq & P_v(\sum_{i=n-k+1}^{n} Z_{i}<cg_0, \;\; 1\leq k\leq n, \; 1\leq n\leq bc-1)\\
 &\geq &\prod_{n=1}^{bc-1}\prod_{k=1}^{ n}(1-\exp\{-cg_0\theta +k\ln h_v(\theta)\})\\
&\geq & [1-\exp\{-cg_0\theta_0\}]^{(bc)^2} \to 1
\end{eqnarray*}
as $c\to +\infty$, where $\theta_0>0$ satisfies $h_v(\theta_0)=1$.

Thus
\begin{eqnarray*}
P_{v}(T_c(g)>m)=P_v(U_m V_m W_{m}S_{bc})\searrow  e^{-t}.
\end{eqnarray*}
as $c\to \infty$. This implies that
\begin{eqnarray*}
 E_{v}(T_c(g)) \geq e^{cg(\mu)\theta^*_v(1+o(1))}/bc
\end{eqnarray*}
for a large $c$. This completes the proof of (A.10).

Let $v\in V^0$.

Let $m_1=(cg(0))^2/\sigma^2$. It follows that
\begin{eqnarray*}
E_v(T_c(g))&=&\sum_{n=0}^{\infty}P_v(T_c(g)> n)\\
&\leq&  m_1+\sum_{n=m_1}^{2m_1}P_v(T_c(g)> n)+...+\sum_{n=km_1}^{(k+1)m_1}P_v(T_c(g)> n)+....\\
&\leq & m_1[1+\sum_{k=1}^{\infty}P_v(T_c(g)> km_1)].
\end{eqnarray*}
Note that
\begin{eqnarray*}
&&P_v(T_c(g)> km_1)\leq  P_v\Big(\sum_{i=(j-1)m_1+1}^{jm_1}Z_{i}+\sum_{i=jm_1-ac+1}^{jm_1}Z'_{i}<cg(0)(1+o(1)), \,\, 1\leq j \leq k \Big)\\
&=& [P_v\Big(\sum_{i=1}^{m_1}Z_{i}+\sum_{i=m_1-ac+1}^{m_1}Z'_{i}<cg(0)(1+o(1))\Big)]^k\\
&=&[P_v\Big(\sum_{i=1}^{m_1-ac}Z_{i}+\sum_{i=m_1-ac+1}^{m_1}(1+A)Z_{i}\,< cg(0)(1+o(1))\Big)]^k\\
&=&[P_v\Big(\frac{\sum_{i=1}^{m_1-ac}Z_{i}}{cg(0)}+\frac{\sum_{i=m_1-ac+1}^{m_1}(1+A)Z_{i}}{cg(0)}\,<
(1+o(1))\Big)]^k
\end{eqnarray*}
for a large $c$, where $A=|g'(0)|/a$,  and
\begin{eqnarray*}
\frac{\sum_{i=1}^{m_1-ac}Z_{i}}{cg(0)} \Rightarrow X\sim N(0,
1),\,\,\,\,\,\,\,\,
\frac{\sum_{i=m_1-ac+1}^{m_1}(1+A)Z_{i}}{cg(0)}\to 0
\end{eqnarray*}
as $c\to \infty$. Thus
\begin{eqnarray*}
E_v(T_c(g))\leq
m_1[1+\sum_{k=1}^{\infty}[\Phi(1+o(1))]^k]=\frac{(cg(0))^2}{\sigma^2}\frac{(1+o(1))}{1-\Phi(1)},
\end{eqnarray*}
where $\Phi(.)$ is the standard normal distribution.

Let $m_2=(cg(0))^2/(8\sigma^2 \ln c)$. Note that
\begin{eqnarray*}
&&\prod_{n=1}^{ac}\prod_{k=1}^nP_v\Big(\sum_{i=n-k+1}^{n}Z_{i}+c|g'(0)|n^{-1}\sum_{i=1}^{n}Z_{i}<cg(0)(1+o(1))\Big)\\
&&\geq
[P_v\Big(\sum_{i=1}^{ac}Z_{i}+\sum_{i=1}^{ac}Z'_{i}<cg(0)(1+o(1))\Big)]^{(ac)^2}=(1+o(1))(1-\Phi(\frac{\sqrt{c}g(0)}{\sqrt{a}(1+A)}))^{(ac)^2}
\to 1
\end{eqnarray*}
as $c\to \infty$, since
\begin{eqnarray*}
P_v(\frac{\sum_{i=1}^{ac}Z_{i}+\sum_{i=1}^{ac}Z'_{i}}{\sigma
(1+A)\sqrt{ac}}) \Rightarrow X \sim N(0, 1)
\end{eqnarray*}
as $c\to \infty$.  It follows  that
\begin{eqnarray*}
&&E_v(T_c(g))\geq \sum_{n=0}^{m_2}P_v(T_c(g)> n)\geq   m_2P_v(T_c(g)> m_2)\\
&\geq & m_2(1+o(1))\prod_{n=ac+1}^{m_1}\prod_{k=1}^nP_v\Big(\sum_{i=n-k+1}^{n}Z_{i}+\sum_{i=n-ac+1}^{n}Z'_{i}<cg(0)(1+o(1))\Big)\\
&\geq &  m_2(1+o(1))[P_v\Big(\sum_{i=1}^{m_2}Z_{i}+\sum_{i=m_2-ac+1}^{m_2}Z'_{i}<cg(0)(1+o(1))\Big)]^{m^2_2}\\
&=&m_2(1+o(1))[P_v\Big(\frac{\sum_{i=1}^{m_2}Z_{i}+\sum_{i=m_2-ac+1}^{m_2}Z'_{i}}{\sqrt{m_2}\sigma}<\sqrt{8\ln c}(1+o(1))\Big)]^{m^2_2}\\
&=&m_2(1+o(1))[\Phi(\sqrt{8 \ln
c})]^{m^2_2}=m_2(1+o(1))[1-\frac{1}{c^4\sqrt{8 \ln c}}]^{m^2_2}\to
m_2(1+o(1))
\end{eqnarray*}
as $c\to \infty$, where the third inequality comes from Theorem 5.1
in Esary, Proschan and Walkup (1967). Thus, we have
\begin{eqnarray*}
\frac{(cg(0))^2}{8\sigma^2 \ln c}(1+o(1))\leq E_v(T_c(g))\leq
\frac{(cg(0))^2}{\sigma^2(1-\Phi(1))}(1+o(1)).
\end{eqnarray*}

Let $v\in V^+$ and let
\begin{eqnarray*}
T_0=\min\{n: \sum_{i=1}^{n}Z_i+\frac{ac}{(ac)\wedge
n}\sum_{i=n-(ac)\wedge n+1}^{n}Z'_{i} \geq c\}.
\end{eqnarray*}
The uniform integrability of $\{T_c(g)/c\}$ for $c\geq 1$, follows
from the well-known uniform integrability of $\{T_0/c\}$ (see Gut
(1988)).

By the Strong Large Number Theorem we have
\begin{eqnarray*}
\mu &=&\lim_{n\to \infty}\frac{\sum_{i=1}^{n}Z_{i}}{n}\\
&=& \lim_{n\to \infty}\max_{1\leq j\leq
n}\frac{1}{n}[\sum_{i=j}^nZ_i+
g'(\mu)a^{-1}\sum_{i=n-ac+1}^{n}(Z_{i}-\mu)]
\end{eqnarray*}
Note that $T_c(g) \to \infty$ as $c\to \infty$,
\begin{eqnarray*}
\max_{1\leq j\leq T_c(g)}[\sum_{i=j}^{T_c(g)}Z_i
+g'(\mu)a^{-1}\sum_{i=T_c(g)-ac+1}^{T_c(g)}(Z_{i}-\mu)]\geq
cg(\mu)(1+o(1)),
\end{eqnarray*}
and
\begin{eqnarray*}
\max_{1\leq j\leq T_c(g)-1}[\sum_{i=j}^{T_c(g)-1}Z_i
+g'(\mu)a^{-1}\sum_{i=T_c(g)-ac}^{T_c(g)-1}(Z_{i}-\mu)]\leq
cg(\mu)(1+o(1)).
\end{eqnarray*}
It follows  that
\begin{eqnarray*}
&&\mu \longleftarrow \max_{1\leq j\leq T_c(g)}\frac{1}{T_c(g)}[\sum_{i=j}^{T_c(g)}Z_i +g'(\mu)a^{-1}\sum_{i=T_c(g)-ac+1}^{T_c(g)}(Z_{i}-\mu)]\\
&\geq & \frac{cg(\mu)(1+o(1))}{ T_c(g)}\\
& \geq  &\max_{1\leq j\leq
T_k-1}\frac{1}{T_c(g)-1}[\sum_{i=j}^{T_c(g)-1}Z_i
+g'(\mu)a^{-1}\sum_{i=T_c(g)-ac}^{T_c(g)-1}(Z_{i}-\mu)]
\longrightarrow \mu
\end{eqnarray*}
as $c\to \infty$. By the  uniform integrability of $\{T_c(g)/c\}$
and using Theorem A.1.1 in  Gut's book (1988), we have
\begin{eqnarray*}
E_v(T_c(g))=(1+o(1))\frac{cg(\mu)}{\mu}
\end{eqnarray*}
for a large $c$. This completes the proof of Theorem 3.

\bigskip

\textbf{Proof of Theorem \ref{theorem4}.} Since $g(x)<0$ for $x>a^*$, $a^*\leq \mu^*$ and $\mu^*\geq 0$, it follows that
\begin{eqnarray*}
P_v\Big( m\hat{Z}_m <  cg(\hat{Z}_m),\,\,\,\hat{Z}_m > a^* \Big)\leq
P_v( \hat{Z}_m<\mu^*)
\end{eqnarray*}
and
\begin{eqnarray*}
P_v(T_c(g)>m)&=&P_v\Big(\sum_{i=n-k+1}^{n}Z_i <  cg(\hat{Z}_n), \,\,\, 1\leq k\leq n, \,\,1\leq n\leq m\Big)\leq P_v\Big(m\hat{Z}_m <  cg(\hat{Z}_m)\Big)\\
&= & P_v\Big(m\hat{Z}_m <  cg(\hat{Z}_m),\,\,\hat{Z}_m \leq  a^* \Big) +P_v\Big( m\hat{Z}_m <  cg(\hat{Z}_m),\,\,\,\hat{Z}_m > a^* \Big)\\
&\leq & 2P_v( \hat{Z}_m<\mu^*).
\end{eqnarray*}
Furthermore,
\begin{eqnarray*}
P_v( \hat{Z}_m<\mu^*)&=&P_v(\sum_{i}^{m}-Z_i >-m\mu^*)=P_v(\sum_{i}^{m}(\mu-Z_i) >m(\mu-\mu^*))\\
 &=& P_v(e^{\theta\sum_{i}^{m}(\mu-Z_i)} >e^{\theta m(\mu-\mu^*)})\leq e^{-m[\theta (\mu-\mu^*)-\ln M(\theta)]},
\end{eqnarray*}
where $M(\theta)=E_v(e^{\theta(\mu-Z_1)})$ and the last inequality
follows from Chebychev's inequality. Note that $h(\theta)=\theta
(\mu-\mu^*)-\ln M(\theta)$ attains its maximum  value
$h(\theta^*)=\theta^*(\mu-\mu^*)-\ln M(\theta^*)>0$ at
$\theta=\theta^*>0$, where $h'(\theta^*)=0$. So,
\begin{eqnarray*}
E_v(T_c(g))=1+\sum_{m=1}^{\infty}P_v(T_c(g)>m)\leq
1+2\sum_{m=1}^{\infty}e^{-m[\theta^* (\mu-\mu^*)-\ln
M(\theta^*)]}=\frac{e^{\theta^* (\mu-\mu^*)-\ln
M(\theta^*)}+1}{e^{\theta^* (\mu-\mu^*)-\ln M(\theta^*)}-1}.
\end{eqnarray*}
Let $k>1$. It follows that
\begin{eqnarray*}
E_{vk}(T_c(g)-k+1)^+&=&\sum_{m=1}^{\infty}P_{vk}(T_c(g)>m+k-1, T_c(g)>k-1 )\\
&\leq & (a_0+1)(k-1)P_0(T_c(g)>
k-1)+\sum_{m\geq(a_0+1)(k-1)}^{\infty}P_{vk}(T_c(g)>m+k-1).
\end{eqnarray*}
Similarly, we have
\begin{eqnarray*}
&&P_{vk}(T_c(g)>m+k-1)\\
&=&P_{vk}\Big(\sum_{i=n-k+1}^{n}Z_i <  cg(\hat{Z}_n), \,\,\, 1\leq k\leq n, \,\,1\leq n\leq m+k-1\Big) \leq  2P_{vk}( \hat{Z}_{m+k-1}<\mu^*)\\
&=&2P_{vk}\Big( \sum_{i=k-1}^{m+k-1}(\mu-Z_i)+ \sum_{i=1}^{k-1}(\mu_0-Z_i)>m(\mu-\mu^*)+(k-1)(\mu_0-\mu^*)\Big)\\
&\leq & 2\exp\{-m\Big(\theta^*(\mu-\mu^*)-\ln
M(\theta^*)+\frac{k-1}{m}[\mu_0-\mu^*-\ln M_0(\theta^*)]\Big)\}\leq
e^{-mb}
\end{eqnarray*}
for $m\geq (a_0+1)(k-1)$, since
\begin{eqnarray*}
\theta^*(\mu-\mu^*)-\ln M(\theta^*)+\frac{k-1}{m}[\mu_0-\mu^*-\ln
M_0(\theta^*)]\geq b
\end{eqnarray*}
for $m\geq (a_0+1)(k-1)$. Thus,
\begin{eqnarray*}
E_{vk}(T_c(g)-k+1)^+&\leq &(a_0+1)(k-1)P_0(T_c(g)\geq k)+2\sum_{m\geq (a_0+1)(k-1)}^{\infty} e^{-mb}\\
&\leq & (a_0+1)(k-1)P_0(T_c(g)>\geq
k)+\frac{2e^{-(a_0+1)(k-1)b}}{1-e^{-b}}.
\end{eqnarray*}

\setcounter{equation}{0}
\bigskip
\renewcommand\theequation{A. \arabic{equation}}


\normalsize
\bigskip

\begin{thebibliography}{}
\bibitem[Dekkers et al.(1989)]{bib13} A. L. M. Dekkers, J. H. J. Einmahl, L. De Haan.(1989). A Moment Estimator for the Index of an Extreme-Value Distribution, Ann.Statist.,17(4), 1833-1855.

\bibitem[Basseville and Nikiforov(1993)]{BN1993} Basseville, M. and Nikiforov, I. (1993). $Detection \,\, of \,\, Abrupt \,\, Changes: \,\,Theory \,\, and \,\,$$ Applications.$ Prentice-Hall, Englewood Cliffs.


\bibitem[Bersimis et al.(2018)]{Bersimis2018} Bersimis, S., Sgora, A. and Psarakis, S. (2018). The application of multivariate statistical process monitoring in non-industrial processes. $Qual. \,\, Reliab. \,\, Engng. \,\, Int.$, 15, 526-549.

\bibitem[Bersimis et al.(2007)]{Bersimis2007} Bersimis,S., Psarakis, S. and Panaretos, J. (2007).  Multivariate statistical process control charts: An Overview. $Qual. \,\, Reliab. \,\, Engng. \,\, Int.$, \textbf{23}, 517-543.


\bibitem[Bowers et al.(2012)]{bib1} Bowers, M. C., Tung, W. W., Gao, J. B..(2012). On the distributions of seasonal river flows: Lognormal or power law $?$, Water Resources Research, 48(5), 0043-1397.

\bibitem[Celano and Castagliola(2018)]{Celano and Castagliola 2018} Celano, G. and Castagliola, P. (2018). An EWMA sign control chart with varying control limits for finite horizon processes. $Qual. \,\, Reliab. \,\, Engng. \,\, Int.$, 34, 1717-C1731

\bibitem[Chakrabortia and Graham(2019)]{CG2019} Chakrabortia,S.and Graham,M.A.(2019). Nonparametric (distribution-free) control charts: An updated overview and some results.  $Qual.\,\, Engineering$, DOI: 10.1080/08982112.2018.1549330

\bibitem[Chatterjee and Qiu(2009)]{Chatterjee and Qiu 2009}  Chatterjee, S. and Qiu, P. (2009). Distribution-free cumulative sum control charts using bootstrap-based control limits.  $Ann. \,\,Appl. \,\, Statistics.$ 3, 349-369.

\bibitem[Cooke, Nieboer, and Misiewicz(2014)]{Coo14} Cooke, R. M., Nieboer, D. and Misiewicz, J. (2014). Fat-Tailed Distributions: Data, Diagnostics and Dependence. John Wiley $\&$ Sons.





\bibitem[Han and Fugee(2004)]{Han and Fugee 2004} Han, D. and Tsung, F. G. (2004). A generalized EWMA control chart and its comparison with the optimal EWMA, CUSUM and GLR schemes.  Ann. Statist., 32, 316-339.

\bibitem[Han and Fugee(2006)]{Han and Fugee 2006} Han, D. and Tsung, F. G. (2006).  A reference-free Cuscore chart for dynamic mean change detection and a unified framework for charting performance comparison.  J. Amer. Statist. Asso., 101, 368-386. 


\bibitem[Huang et al.(2016)]{Huang2016} Huang, W. P., Shu, L. J., Woodall, W. H. and Tsui, K. L. (2016). CUSUM procedures with probability control limits for monitoring processes with variable sample sizes. $ IIE \,\, Trans$. 48, 759-771.





\bibitem[Lai(2001)]{Lai2001} Lai, T. L. (2001). Sequential analysis: some classical problems and new challenges. $Statist \,\, Sinica$,  \textbf{11}, 303-408.

\bibitem[Lai(1995)]{Lai1995}  Lai, T. L. (1995). Sequential change-point detection in quality control and dynamical systems. $J.\,\,R.\,\, Statist\,\,Soc.\,\, \textbf{B}.$ 57, 613-658.

\bibitem[Lai(1998)]{Lai1998}  Lai, T. L. (1998). Information Bounds and Quick Detection of Parameter Changes in Stochastic Systems. \emph{IEEE Trans. Inf. Theory}, \textbf{44}, 2917-2929.

\bibitem[Lee et al.(2013)]{Lee2013}   Lee, P. H., Huang, Y. H., Kuo, T. I. and Wang, C. C. (2013). The effect of the individual chart with variable control limits on the river pollution monitoring. $Qual \,\, Quant$, 47, 1803-1812.

\bibitem[Margavio et al.(1995)]{Margavio1995} Margavio, T. M., Conerly, M. D., Woodall, W. H. and Drake, L. G. (1995). Alarm rates for quality control charts. $Stat.\,\, Probab.\,\,  Lett. $, 24, 219-C224

\bibitem[Montgomery(2009)]{Montgomery2009} Montgomery, D. C. (2009). $Introduction \,\, to \,\, Statistical \,\, Quality \,\,  Control$. 6th ed. New York: John Wiley \& Sons.

\bibitem[Moustakides(1986)]{Moustakides1986} Moustakides, G. V. (1986). Optimal stopping times for detecting changes in distributions. Ann. Statist., 14, 1379–1387.

\bibitem[Page(1954)]{Page 1954} Page, E. S. (1954). Continuous inspection schemes,$Biometrika$ 41, 100-115.



\bibitem[Poor and Hadjiliadis(2008)]{Poor2008} Poor, H. V., \& Hadjiliadis, O. (2008). Quickest detection. Cambridge University Press. https://doi.org/10.1017/CBO9780511754678

\bibitem[Qiu(2014)]{Qiu2014} Qiu, P. (2014). $Introduction\,\, to \,\, Statistical \,\, Process\,\,  Control$. Boca Raton, FL: Chapman and Hall/CRC.

\bibitem[Resnick(1989)]{bib3} Resnick, S. I..(1989). Extreme values, regular variation, and point processes, Journal of the American Statistical Association, 84(407), 845.

\bibitem[Resnick(2007)]{bib4} Resnick, S. I..(2007). Heavy-Tail Phenomena: Probabilistic and Statistical Modeling, Springer Verlag, Springer Series in Operations Research and Financial Engineering, 2007.

\bibitem[Ritov(1990)]{Ritov1990} Ritov, Yaacov.(1990). Estimation in a Linear Regression Model with Censored Data .Ann. Statist., 18, 303-328.


\bibitem[Shen et al.(2013)]{Shen2013} Shen, X., Zou, C. L., Jiang, W. and Tsung, F. G. (2013) Monitoring Poisson count

\bibitem[Shewhart(1931)]{Shewhart1931} Shewhart, W. A. (1931)  $Economic \,\, Control \,\, of \,\, Quality \,\, of \,\, Manufactured \,\, Product$. New York: Van Nostrand.

\bibitem[Siegmund(1985)]{SD1995} Siegmund, D. (1985). $Sequential\,\, Analysis: \,\,Tests\,\, and \,\,Confidence\,\, Intervals$. Springer, New York.

\bibitem[Sogandi et al.(2019)]{Sogandi2019} Sogandi, F., Aminnayeri, M.,  Mohammadpour, A. and Amiri, A. (2019)  Risk-adjusted Bernoulli chart in multi-stagehealthcare processesbased onstate-space modelwith alatent riskvariable and dynamic probability control limits. $Comp.\, \, Industr. \,\, Engin.$ 130, 699-713

\bibitem[Steiner(1999)]{Steiner 1999}  Steiner, S. H. (1999). EWMA control charts with time-varying control limits and fast initial response. $J. \,\, Quality \,\, Technology$, 31, 75-86.

\bibitem[Stoumbos et al.(2000)]{Stoumbos2000} Stoumbos, Z. G., Reynolds, M. R., Ryan, T. P., and Woodall, W. H. (2000) The state of statistical process control as we proceed into the 21st century. $J. \,\,Amer. \,\,Statist. \,\,  Assoc.$, \textbf{95}, 992-998.

\bibitem[Tang and Han(2023)]{tang2023}Fuquan Tang and Dong Han. (2023). The asymptotic distribution of a truncated sample mean for the extremely heavy-tailed distributions, Communications in Statistics - Theory and Methods, DOI: 10.1080/03610926.2023.2209231

\bibitem[Verdier et al.(2017)]{Verdier2017}  Verdier, G., Hilgert, N. and Vila, J. P. (2008). Adaptive threshold computation for CUSUM-type procedures in change detection and isolation problems.   $Computational  \,\,Statistics.\,\, \& \,\, Data \,\, Analysis$, 52, 4161-4171.

\bibitem[Woodall et al.(2017)]{Woodall2017} Woodall, W. H., Zhao, M. J., Paynabar, K., Sparks, R. and Wilson, J. D. (2017)  An overview and perspective on social network monitoring. $IIE \,\, Trans.$, \textbf{49}, 354-365.


\bibitem[Yang et al.(2017)]{Yang2017}    Yang, W. W., Zou, C. L. and Wang, Z. J. (2017). Nonparametric profile monitoring using dynamic probability control limits. $Qual. \,\, Reliab. \,\, Engng. \,\, Int.$, 33,  1131-C1142

\bibitem[Zhang and Woodal(2015)]{Zhang and Woodal 2015}  Zhang, X. and Woodal, W. H. (2015). Dynamic probability control limits for risk-adjusted Bernoulli CUSUM chart. $Statist.\,\, Med.$, 34, 3336-C3348

\bibitem[Zhang and Woodal(2017)]{Zhang and Woodal 2017}  Zhang, X. and Woodal, W. H. (2017). Reduction of the effect of estimation error on in-control performance for risk-adjusted
Bernoulli CUSUM chart with dynamic probability control limits. $Qual. \,\, Reliab. \,\, Engng. \,\, Int.$, 33,  381-C386



























\end{thebibliography}
\end{document}